# Microstrip Patch Antenna Simulation using Comsol Multiphysics

R. P. Gawade, S. G. Dahotre*

*Department of Physics, Dr. Babasaheb Ambedkar Technological University, Lonere-402103, Raigad, India*

gawade.rohini95@gmail.com, sgdahotre@dbatu.ac.in

**Abstract:**

In this renovating era of technology, human life has considerably changed like transport has been transformed from bull cart to aircraft, land line phones to 4G/5G cellular phones, CRT based T. V to LED T. V etc. Human being has perceived the huge revolution especially in the field of communication. We report on the simulation study of Microstrip patch antenna using Comsol Multiphysics 5.4 software. Transmitter, feed point, receiver, modulation are the several parameters which are most important during the fabrication of antenna. In this paper we explain the geometry, radiation patterns in E plane and H plane, scattering parameters, resonant frequency, bandwidth of microstrip patch antenna. In research field lot of work is done on antenna by many of the scientists but there is a challenge about the compactness of antenna. Microstrip patch antenna is much compact than other antennas and these types of antennas are used in our mobile. It attains the greater number of frequency bands, improved gain, better impedance matching and improved bandwidth. In this paper, frequency response of Microstrip Patch Antenna shows single band behaviour at frequency 1.575 GHz which gives improvement in reflection coefficient and wider bandwidth.

**Keywords:** Far-field domain, Metamaterial, impedance, polarization, resonant frequency, Comsol Multiphysics.

**Introduction:**

In modern communication system antenna plays vital role. There are different types of antennas such as, Monopole, Dipole, Yagi-Huda, Conical, Lens, urf-rfid, Microstrip Patch antenna etc. In this paper we review some of the most noteworthy advances in the design and demonstrating of microstrip antennas that have been fabricated in the last several years. The idea of the microstrip antenna dates back to the 1950's but it was not until the 1970's that serious attention was given to this element. A microstrip antenna contains metallic patch printed on a thin, grounded dielectric substrate. The microstrip antenna has a very compact, and can be invented using printed circuit (3D printing, photolithography) techniques [1]. This implies that the antenna can be made accordant, and possibly at low cost. Other advantages are like easy fabrication into linear or planar arrays, and easy integration with microwave integrated circuits, improved bandwidth, improved gain, better impedance matching, improved efficiency etc [2-4]. Additional parameters of this antenna are patch shape, feeding techniques, substrate configurations, and array geometries which have been invented by researchers throughout the world [5,6]. To a large extent, the development of microstrip antennas has been driven by systems requirements for antennas with low-profile, low-weight, low cost, easy integrability into arrays or with microwave integrated circuits, or polarization diversity [7-9]. Thus, microstrip antennas found application in both the military and the civil sectors, such as aircraft, missiles, satellites, ships, land vehicles, biomedical systems etc. All these applications are achieved when we use metamaterials in the fabrication of microstrip antenna [10-14].

At the first time, synthesis of metamaterial was done by Rodger M. Walser, University of Texas at Austin, in 1999. Metamaterial is a material which is not found in nature but can be prepared artificially. Most of the properties of metamaterials are due to specific shape, size and geometry of the substrate in which we insert our material. The metamaterial is defined as "a material which gains its properties from its structure rather than directly from its composition." These material shows negative refractive index so that it is also called Left-Handed Materials (LHM). In LHM, permittivity and permeability of the material both are negative [15].

**Flow Chart of Computational Work:**

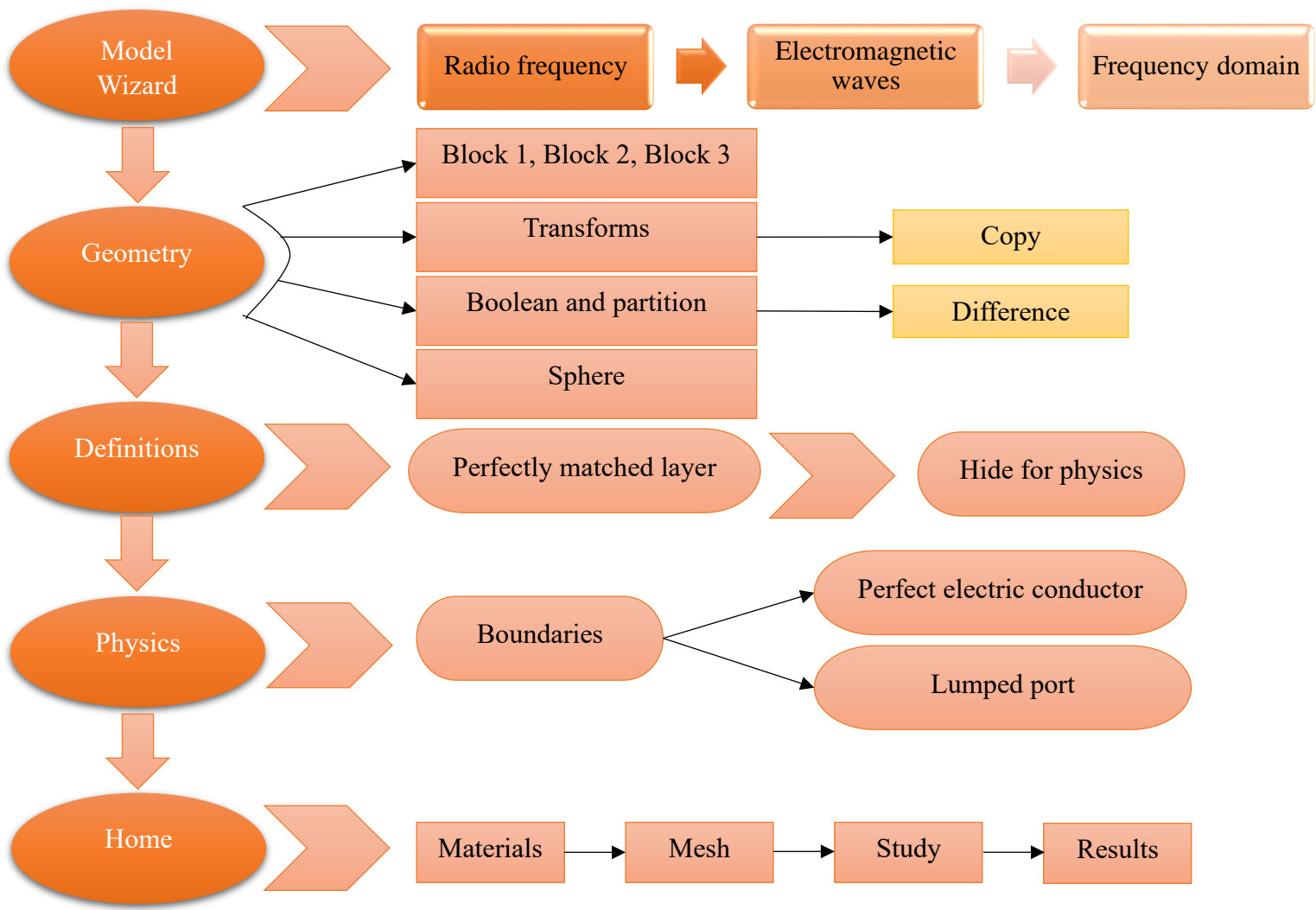

**Fig. 1: Flow chart of computational work**

As shown in Figure 1, initially we have selected 3D model of microstrip patch antenna. All the required parameters which are used for constructing proposed antenna is given in the table 1. Using geometry tool and parameters from Table 1 we create the substrate block which is shown in Figure 2.

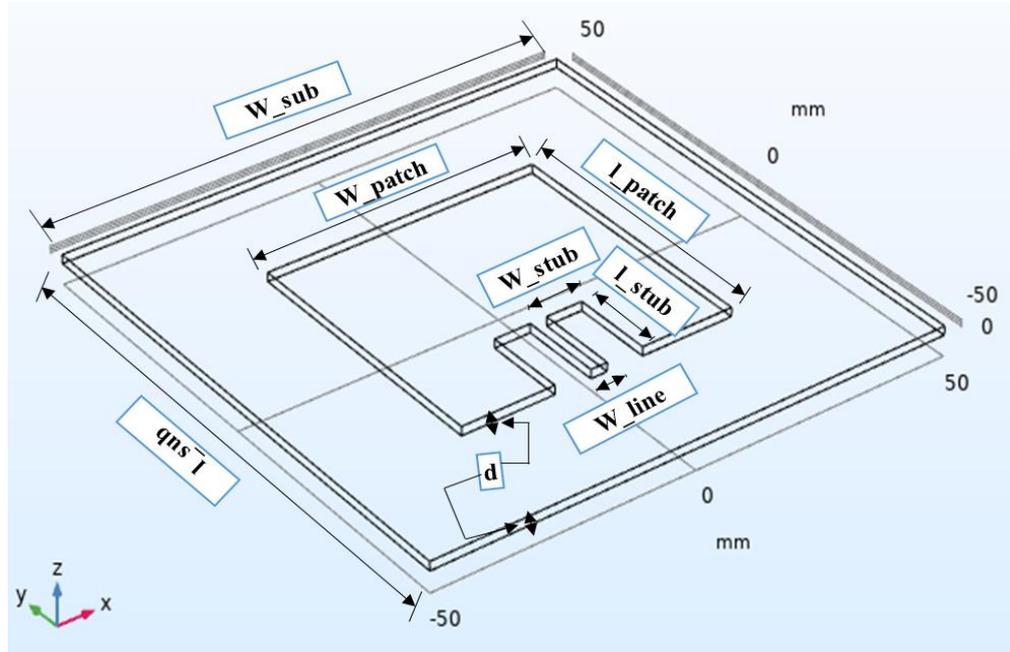

**Fig. 2: Geometry of microstrip patch antenna**

**Table 1: Parameters for making geometry of microstrip patch antenna:**

| Name | Expression | Value | Description |
|---|---|---|---|
| d | 60[mil] | 0.001524m | Substrate thickness |
| w_line | 3.2[mm] | 0.0032m | 50 ohm line width |
| w_patch | 53[mm] | 0.053m | Patch width |
| l_patch | 52[mm] | 0.052m | Patch length |
| w_stub | 7[mm] | 0.007m | Tuning stub width |
| l_stub | 16[mm] | 0.016m | Tuning stub length |
| w_sub | 100[mm] | 0.1m | Substrate width |
| l_sub | 100[mm] | 0.1m | Substrate length |

Here mil refers to the unit milli inch, that is 1 mil = 0.0254 mm.

As shown in Figure 2, w_line added the patch of antenna and created impedance matching parts as well as a 50 Ω feed line. Using sub-tools such as copy, Boolean and partition we get required geometry of proposed antenna. At the same time using wireframe rendering tool we observed the better view of interior parts of substrate. By using wireframe rendering tool we got the better view of interior parts. For the geometry on substrate, some domains and boundary conditions of electromagnetic wave were specified using far-field domain sub-tool within physics tool for getting radiation pattern in E and H- plane. Before adding the material in the geometry specify all physics which was necessary for obtaining the results. The values of relative permittivity and relative permeability of the metamaterial which is added in the geometry are 3.38 F/m and 1 H/m respectively and these are calculated by using Nicolson-Ross-Weir method [16,17]. Thus, we get the required results for microstrip patch antenna using Comsol Multiphysics 5.4 software.

**Results and Discussion from Simulation Study:**

All the simulations of proposed antenna structure have been carried out by using Comsol Multiphysics 5.4 software.

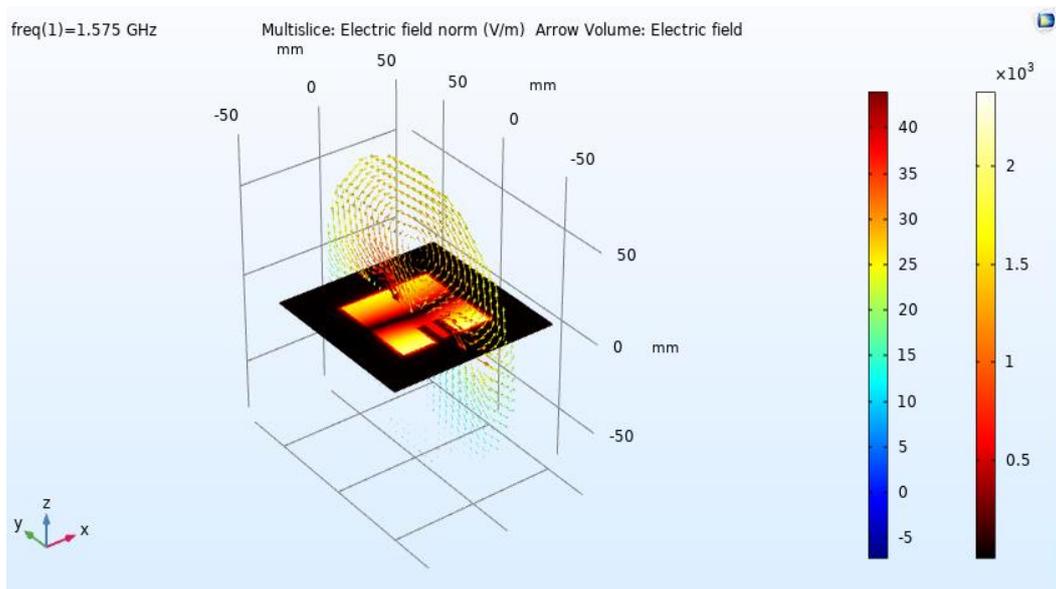

**Fig. 3: Distribution of electric field within the geometry of antenna.**

The norm of electric field inside the antenna substrate is described in Figure 3, where an arrow plot of the electric field is included. The direction of the arrows indicate the dominant polarization in the direction of maximum radiation. The antenna impedance is higher if 50 Ω fed from the edge and lower if 50 Ω fed from the center. Therefore, an optimum feed point exists between the center and the edge.

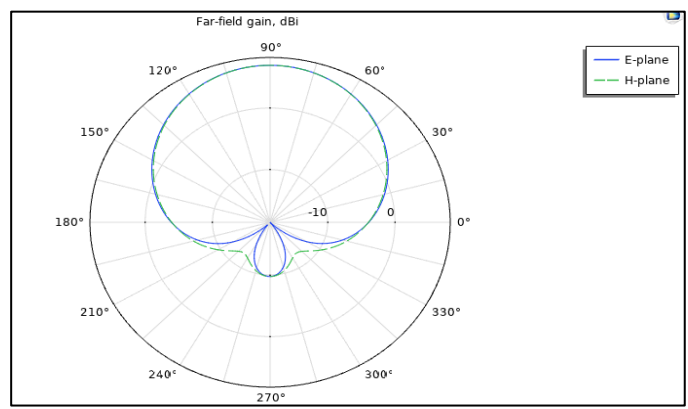

**Fig. 4: Far-field radiation pattern (gain in dBi) at E-plane and H-plane.**

Figure 4 shows the simulated radiation plot of E and H-plane in terms of far-field gain (dBi) at resonant frequency of microstrip patch antenna. Proposed antenna produces almost super cardioid polar pattern because of the bottom ground plane, the radiation pattern is directed toward the top. From that graph it is clear that, proposed antenna is used for enhancing the gain and directional capabilities.

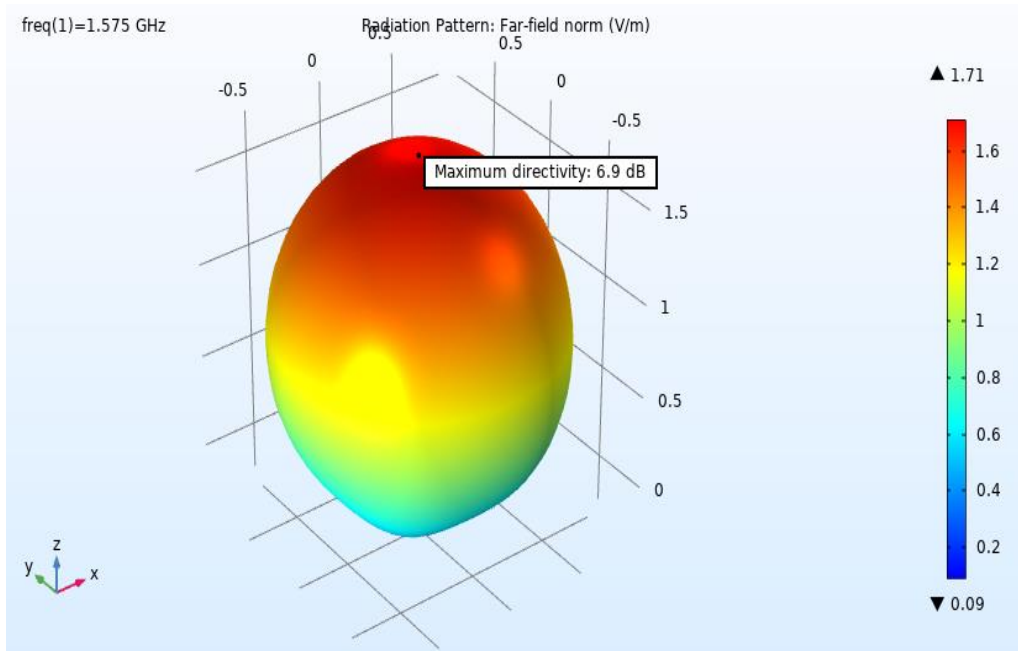

**Fig. 5: 3D Far-field radiation pattern**

Figure 5 shows the directivity of radiation which is evaluated from 3D far-field pattern. The directive beam pattern due to the ground plane that blocks the radiation toward the bottom side. Maximum directivity calculated by using following formula [18]:

$$D = \frac{Maximum\ power\ gain\ by\ antenna}{Average\ power\ gain}$$

$$= \frac{4\pi f(\theta,\phi)}{P{avg}}$$

where, **f** = Resonant frequency (1.575 GHz from Figure 7) and **P**$_{avg}$ = Average power gain (2.85 from Table 1 and Figure 4).

$$= \frac{4 \times 3.14 \times 1.575}{28.5 \times 0.1}$$

$$= \frac{19.782}{2.85}$$

$$\boxed{D = 6.9410\ dB}$$

Thus, the calculated antenna directivity is greater than 6.9 dB.

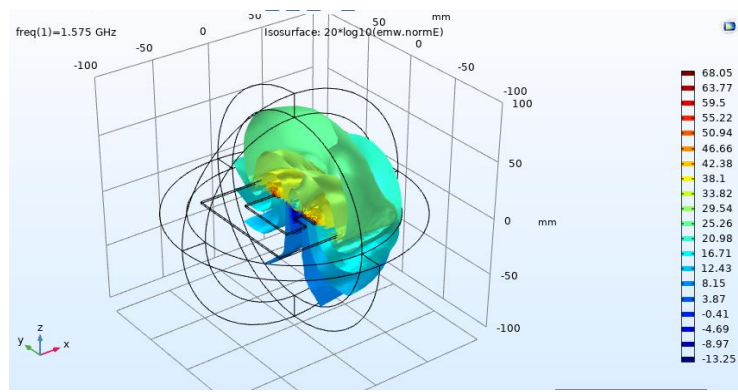

**Fig. 6: An iso-surface plot visualizes the decay of the field amplitude.**

Figure 6 visualizes the decay of field amplitude within proposed geometry of antenna at resonant frequency 1.575 GHz. It shows that variation in number of electric field lines per unit volume within our proposed geometry having values varies from -13.25 to 68.05.

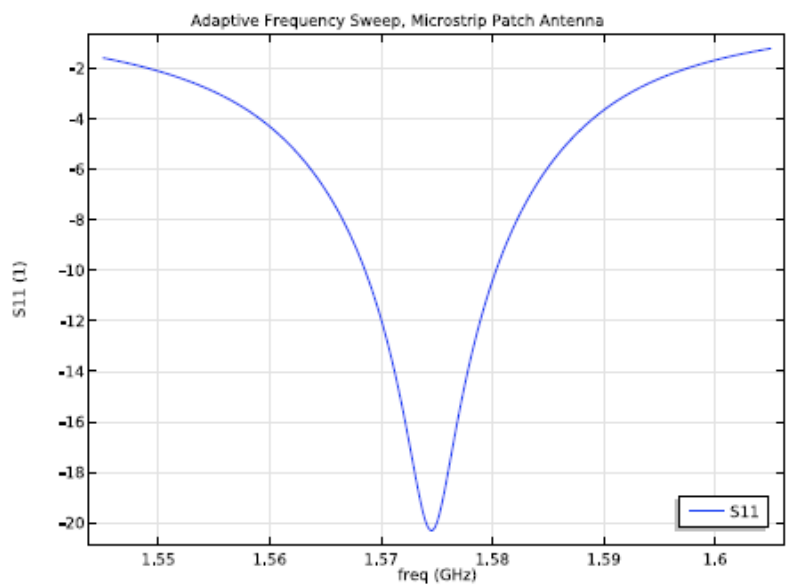

**Fig. 7: Resonant frequency**

The scattering parameter ($S_{11}$) is better than -10 dB, and the front-to-back ratio in the radiation pattern is more than 15 dB. The frequency response is evaluated with 100 kHz resolution which is plotted in Figure 7. It shows that the microstrip patch antenna exhibits single band behaviour and reports only one resonant frequency band at 1.575 GHz. By observing the resonance point of proposed antenna, it can be predicted that the frequency range of proposed antenna lies between 1.55 GHz to 1.60 GHz which indicates improvement in reflection coefficient and wider bandwidth at respective point of resonance. S-parameter (S11) plot shows that the antenna impedance is matched to 50 Ω around 1.575 GHz.

The operators used for the far field radiation pattern of a virtual 8x8 microstrip patch antenna array are shown in Table 2.

**Table 2: Input arguments of array factor operator for an 8x8 array:**

| ARGUMENT | DESCRIPTION | ARGUMENT | UNIT |
|---|---|---|---|
| nx | Number of elements along x-axis | 8 | Dimensionless |
| ny | Number of elements along y-axis | 8 | Dimensionless |
| nz | Number of elements along z-axis | 8 | Dimensionless |
| dx | Distance between array elements along x-axis | 0.48 | Wavelength |
| dy | Distance between array elements along y-axis | 0.48 | Wavelength |
| dz | Distance between array elements along z-axis | 0 | Wavelength |
| alphax | Phase progression along x-axis | 0 | Radian |
| alphay | Phase progression along y-axis | 0 | Radian |
| alphaz | Phase progression along z-axis | 0 | Radian |

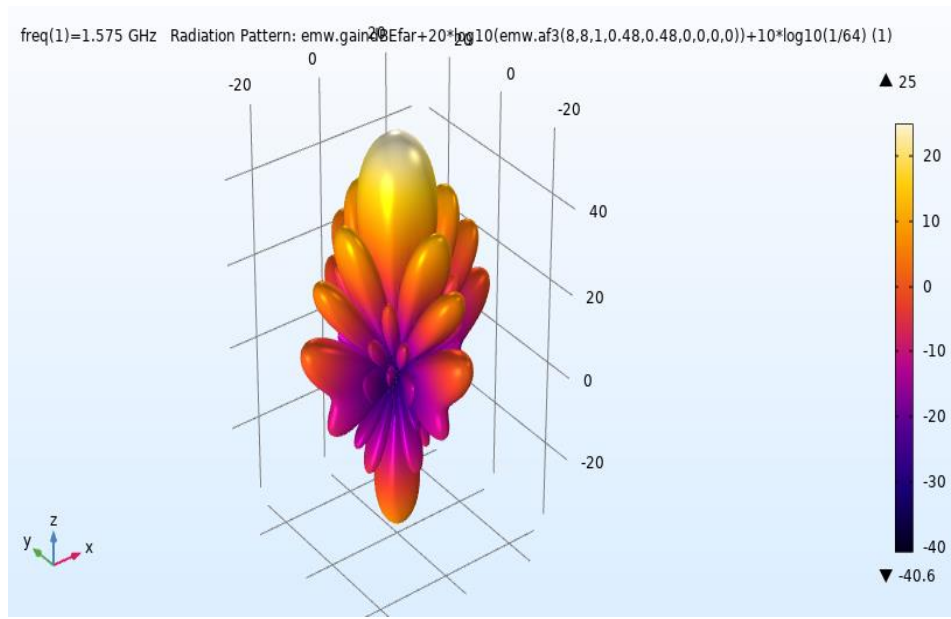

**Fig. 8: The far-field radiation pattern of a virtual 8×8 microstrip patch antenna array.**

Figure 8 shows that radiation pattern of 8x8 microstrip patch antenna array. As shown in table 2 we can construct the array of eight antenna elements along x, y, z axis. Radiation pattern of this array shows the variation in the radiations along all three axes. The maximum radiation direction of the array factor along the x-axis is defined by angle θ, along the y-axis is defined by the radial distance r and along the z-axis is defined by angle ϕ.

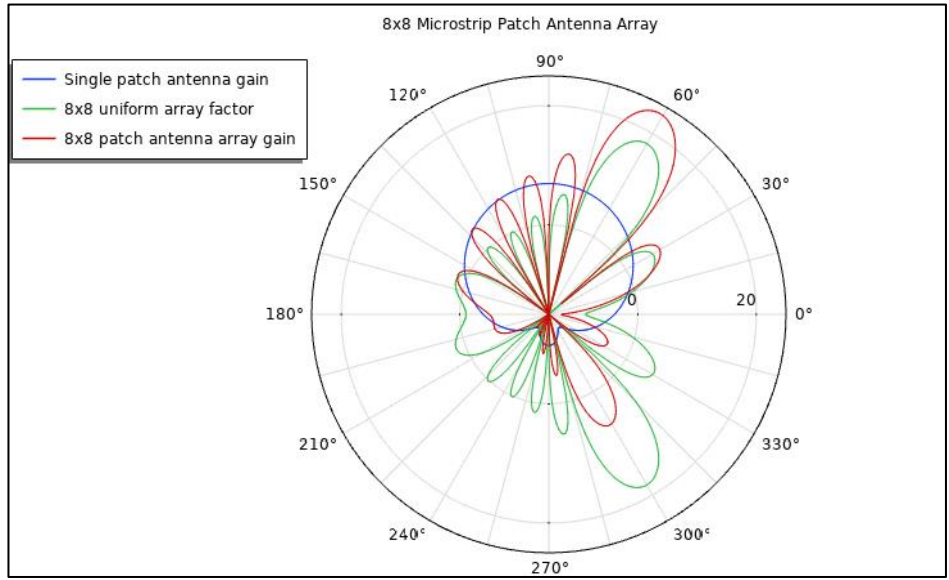

**Fig. 9: The single patch antenna gain, 8×8 uniform array factor, 8×8 microstrip patch antenna array gain plotted in dB-scale.**

Figure 9 includes three plots to show the evolution of the antenna radiation pattern from a single antenna to a synthesized antenna array via the uniform array factor. This shows that single patch antenna gain, position of antenna elements and gain of 8x8 microstrip patch antenna array. The radiation pattern of the uniform array factor for 8x8 microstrip patch antenna configured to have the maximum radiation at 60 degrees.

**Conclusions:**

The geometry of proposed antenna is much compact than other antennas. Using rectangular microstrip patch antenna design we get better polarization at the edge than the centre of geometry of proposed antenna. The radiation pattern of single patch antenna in E and H-plane, 8x8 microstrip patch antenna, which shows the directivity and gain improvement of antenna. Resonant frequency of proposed antenna is 1.575GHz which is applicable for bandwidth enhancement and better impedance matching. Thus, efficiency of microstrip patch antenna is high. So, we can say that such type of antenna is useful for wireless communication systems like, mobile, Wi-Fi, Television and radio broadcasting, satellite communication and Radar system.

**Acknowledgement:**

We acknowledge the Dr. Babasaheb Ambedkar Technological University, Lonere for their support. The authors acknowledge Dr. P. S. Alegaonkar, Associate Professor and Head of department of Physics, Central University, Panjab and Mr. S. L. Chinke, Assistant Professor, Savitribai Phule Pune University, Pune for their guidance.